\documentclass[twocolumn]{aastex63}
\usepackage[all]{hypcap}

\graphicspath{{./}{figures/}}

\begin{document}

\title{Determining the nanoflare heating frequency of an X-ray Bright Point observed by MaGIXS}

\correspondingauthor{Biswajit Mondal}
\email{biswajit.mondal@nasa.gov, biswajit70mondal94@gmail.com}

\author[0000-0002-7020-2826]{Biswajit Mondal}
\affiliation{NASA Postdoctoral Program, NASA Marshall Space Flight Center, ST13, Huntsville, AL, USA}
\author[0000-0002-4454-147X]{P. S. Athiray}
\affiliation{Center for Space Plasma and Aeronomic Research, The University of Alabama in Huntsville, Huntsville, AL, USA}
\affiliation{NASA Marshall Space Flight Center, ST13, Huntsville, AL, USA}
\author[0000-0002-5608-531X]{Amy R. Winebarger}
\affiliation{NASA Marshall Space Flight Center, ST13, Huntsville, AL, USA}
\author{Sabrina L. Savage}
\affiliation{NASA Marshall Space Flight Center, ST13, Huntsville, AL, USA}
\author{Ken Kobayashi}
\affiliation{NASA Marshall Space Flight Center, ST13, Huntsville, AL, USA}
\author{Stephen Bradshaw}
\affiliation{Department of Physics and Astronomy (MS 108), Rice University, 6100 Main Street, Houston, TX 77005, USA}
\author[0000-0001-9642-6089]{Will Barnes}
\affiliation{Department of Physics, American University, Washington, DC, USA}
\affiliation{NASA Goddard Space Flight Center, Heliophysics Science Division, Greenbelt, MD 20771, USA}
\author{Patrick R. Champey}
\affiliation{NASA Marshall Space Flight Center, ST13, Huntsville, AL, USA}
\author{Peter Cheimets}
\affiliation{Center for Astrophysics | Harvard \& Smithsonian, Cambridge, MA 02138, USA}
\author{Jaroslav Dudík}
\affiliation{Astronomical Institute,Czech Academy of Sciences, Fricova 298,25165 Ondrejov, CzechRepublic}
\author{Leon Golub}
\affiliation{Center for Astrophysics | Harvard \& Smithsonian, Cambridge, MA 02138, USA}
\author[0000-0002-6418-7914]{Helen E. Mason}
\affiliation{DAMTP, Centre for Mathematical Sciences, University of Cambridge, Wilberforce Road, Cambridge CB3 0WA, UK}
\author{David E. McKenzie}
\affiliation{NASA Marshall Space Flight Center, ST13, Huntsville, AL, USA}
\author{Christopher S. Moore}
\affiliation{Center for Astrophysics | Harvard \& Smithsonian, Cambridge, MA 02138, USA}
\author{Chad Madsen}
\affiliation{Center for Astrophysics | Harvard \& Smithsonian, Cambridge, MA 02138, USA}
\author{Katharine K. Reeves}
\affiliation{Center for Astrophysics | Harvard \& Smithsonian, Cambridge, MA 02138, USA}
\author[0000-0002-0405-0668]{Paola Testa}
\affiliation{Center for Astrophysics | Harvard \& Smithsonian, Cambridge, MA 02138, USA}
\author{Genevieve D. Vigil}
\affiliation{NASA Marshall Space Flight Center, ST13, Huntsville, AL, USA}
\author{Harry P. Warren}
\affiliation{Space Science Division, Naval Research Laboratory, Washington, DC 20375, USA}
\author{Robert W. Walsh}
\affiliation{University of Central Lancashire, Preston, PR1 2HE, UK}
\author[0000-0002-4125-0204]{Giulio Del Zanna}
\affiliation{DAMTP, Centre for Mathematical Sciences, University of Cambridge, Wilberforce Road, Cambridge CB3 0WA, UK}

\begin{abstract}
Nanoflares are thought to be one of the prime candidates that can heat the solar corona to its multi-million kelvin temperature. Individual nanoflares are difficult to detect with the present generation instruments, however their presence can be inferred by comparing simulated nanoflare-heated plasma emissions with the observed emission. 
Using HYDRAD coronal loop simulations, we model the emission from an X-ray bright point (XBP) observed by the Marshall Grazing Incidence X-ray Spectrometer (MaGIXS), along with nearest-available observations from the Atmospheric Imaging Assembly (AIA) onboard Solar Dynamics Observatory (SDO) and X-Ray Telescope (XRT) onboard Hinode observatory.
The length and magnetic field strength of the coronal loops are derived from the linear-force-free extrapolation of the observed photospheric magnetogram by Helioseismic and Magnetic Imager (HMI) onboard SDO. Each loop is assumed to be heated by random nanoflares, whose magnitude and frequency are determined by the loop length and magnetic field strength.
The simulation results are then compared and matched against the measured intensity from AIA, XRT, and MaGIXS. Our model results indicate the observed emissions from the XBP under study could be well matched by a distribution of nanoflares with average delay times { 1500~s to 3000~s,} which suggest that the heating is dominated by high-frequency events. Further,  we demonstrate the high sensitivity of MaGIXS and XRT to diagnose the heating frequency using this method, while AIA passbands are found to be  the least sensitive.

\end{abstract}

\section{Introduction}

Understanding the heating of the non-flaring solar corona is an active topic of research in heliophysics.
It is well accepted that magnetic fields are mainly responsible for coronal heating.
The photospheric driver randomly moves the foot-points of the magnetic field lines, and either generates waves or the quasi-static buildup of magnetic energy, depending on the timescale of motion~\citep{Klimchuck_2006SoPh}.
Heating by the dissipation of the magnetic energy (e.g.,~\citealp{parker_1988}) is termed as DC heating while the dissipation of wave (e.g.,~\citealp{Alfven_1947}) is known as AC heating mechanism.
Both the AC and DC heating mechanisms can lead to impulsive heating events, termed nanoflares~\citep{klimchuk_2015RSPTA}.
The magnitude and frequency of these nanoflares determine whether they can adequately satisfy the coronal heating budget. 
Thus it is of great importance to study the nanoflares and determine their frequency to validate their role in coronal heating.
According to the occurrence frequency, the nanoflares are primarily classified into two different categories, High-Frequency (HF) nanoflares, and Low-Frequency  (LF) nanoflares.  HF nanoflares are when the cooling time scale is short compared to the time between two successive heating events.
The plasma 
could not cool enough in between the events, and in this case, the plasma would be heated quasi-steadily. 
On the other hand, for the LF nanoflares heating, the plasma would be cooled significantly before successive events. 
Determining the properties of the nanoflares from the observations would significantly constrain the properties of the heating mechanism.

Due to the small, faint, and transient nature of the nanoflares, their direct observation by the current generation instruments is limited by several factors, including inadequate instrumental spatial resolution, cadence, and spectral information. 
To infer and validate the nanoflare heating scenario, in the absence of direct observations, researchers often used different plasma diagnostics, e.g., emission measure (EM) distribution~\citep{Reale_2009ApJ, Tripathi_2011,Testa_2011ApJ, Warren_2011ApJ,Warren_2012ApJ,Winebarger_2011, Testa_2012ApJ...750L..10T, giulio_2015A&A,Brosius_2014ApJ, Caspi_2015,Ishikawa_2017NatAs}, variability of footpoint emission~\citep{Testa_2013ApJ...770L...1T,Testa_2014Sci...346B.315T} and time-lag analysis~\citep{Viall_2012ApJ,Viall_2017ApJ...842..108V}.
The EM distribution, which indicates the amount of emitting plasma at different temperatures, is a useful diagnostic for parameterizing the frequency of energy deposition. 
Several observational and theoretical studies (e.g., \citealp{Jordan_1976,cargill_1994ApJ,Cargill_2004,Warren_2012ApJ})
have suggested that 
EM has a peak at an average plasma temperature  (for AR, 3-4 MK) along with cool and hot components. 

For a better understanding of the frequency and observable properties of nanoflare heating, several earlier studies compare the observed intensities, EM distribution and/or other observable quantities with the simulated nanoflare heated plasma. 
For example, \cite{Barnes_2019} and \cite{Barnes_2021ApJ...919..132B} compared the EM distribution and time lags of simulated nanoflares heated plasma of AR with the observed distribution derived from EUV observation by the Atmospheric Imaging Assembly (AIA:~\citealp{Lemen_2012SoPh}) on
board SDO~\citep{Pesnell_2012SoPh}. Their study suggests that high-frequency nanoflares dominate the core of the AR. \cite{Warren_2020ApJ...896...51W} compare the modeled EM of an AR with the derived EM from the EUV observations of High-resolution Coronal Imager (Hi-C) sounding rocket experiment. They also found that high-frequency heating provides the best match to the observed EM. 
Recently \cite{mondal_2023} studied the average nanoflare frequency to heat coronal X-ray Bright Points (XBP) by comparing the simulated EM distribution with the observed distribution.
For accurate estimation of the observed EM distribution at higher temperatures, they combine the EUV observations of SDO/AIA with the moderate energy resolution disk-integrated X-ray spectra observed by Solar X-ray Monitor (XSM:~\citealp{xsm_flight_performance,XSM_ground_calibration,xsm_XBP_abundance_2021}) onboard Chandrayaan-2. They found a good match of observed EM distribution at coronal temperatures with the simulated distribution of nanoflare heated plasma. These nanoflares had multiple frequencies, and their energy distribution followed a power-law slope close to -2.5. 
However, as XSM provides the disk-integrated spectrum, it is not efficient for deriving the EM distribution for a single coronal feature (e.g., single AR or XBP). Studying the heating frequency in great detail for a single AR or XBP requires sensitive spatially resolved spectroscopic observation in EUV and X-ray energies.

The Marshall Grazing Incidence Spectrometer (MaGIXS:~\citealp{Champey_2022JAI}) is primarily designed for diagnostics of coronal heating frequency for AR { (see Section~\ref{sec-observation})}.
MaGIXS is a sounding rocket mission whose first successful flight was carried out on 30th July 2021.
In this work, we have studied the heating frequency of an X-ray Bright Point (XBP) that MaGIXS observed.

We have derived the loop structures of the XBP using the potential field extrapolation of the photospheric line-of-sight (LOS) magnetogram observed by the Helioseismic and Magnetic Imager (HMI:~\citealp{scherre_2012SoPh}) onboard the SDO. 
The emission of these XBP loops is simulated using the HYDrodynamics and RADiation code (HYDRAD\footnote{\url{https://github.com/rice-solar-physics/HYDRAD}}:~\citealp{Bradshaw_2003A&A,Bradshaw_2013ApJ...770...12B}) for nanoflare heating, whose frequencies and magnitudes are estimated from the loop parameters (e.g., lengths and magnetic field strengths).
Here, we considered that nanoflares originated from the dissipation of magnetic energy stored within the loop. 
From the simulated outputs, we calculate EM distributions and generate the synthetic images of the XBP. These synthetic images are compared with the MaGIXS observation as well as the { nearest available} EUV and X-ray images observed by the SDO/AIA and the X-Ray Telescope (XRT:~\citealp{Golub_2007SoPh}) onboard Hinode~\citep{Kosugi_2007SoPh}.

Our goal here is to study whether nanoflare heating can explain the observed emission properties of the XBP and to investigate the importance of the MaGIXS, AIA, and XRT observations in determining the frequency of nanoflare heating. 
The rest of the paper is organized as follows. Section~\ref{sec-observation} describes the MaGIXS, AIA, XRT, and HMI observations of the XBP. Section~\ref{sec-simulations} described the simulation setup. The results are shown and discussed in Section~\ref{sec-results}, and Section~\ref{sec-summary} provides a brief summary of the work.

\section{Observations}\label{sec-observation}

{ MaGIXS is a grazing incidence wide-field slot imaging spectrometer, consisting of a Wolter-I telescope, a slot, grating spectrometer, CCD camera, and a slitjaw context imager. It's FOV is restricted by the slot of 12$'$-wide and 33$'$-long (\citealp{Champey_2022JAI}).
The unique design of MaGIXS is optimized to capture a spectral and spatial overlappogram of a solar AR in the soft X-ray wavelength range from $\sim$ 8\,\AA\ to 30\,\AA\, (0.4-1.5 keV).} 
Spectral measurements in this energy range are well suited to diagnose the heating of ARs~\citep{Athiray_2019ApJ}.
The first successful rocket flight of MaGIXS on July 30 2021 at 18:20 UT, was targeted to observe two ARs (12846 and 12849) in the  northern and southern solar hemispheres. { However, due to the 
internal vignetting, the effective FOV observed is 9.2\arcmin $\times$ 25\arcmin on the solar disk, sampling two X-ray bright points (XBP-1, XBP-2) and a portion of the active region (AR 12849)} (see \citealp{Sabrina_2023ApJ} for the details of MaGIXS observation).
It recorded 296 s of imaging spectroscopic observations with a cadence of 2 s. { For this analysis, we utilized MaGIXS Level 2 data products, which are spectrally pure images of the X-ray bright points (see table~3 of \citealp{Sabrina_2023ApJ} and text for a description of the data processing).}  

\begin{figure*}[ht!]
\centering
\includegraphics[width=1\linewidth]{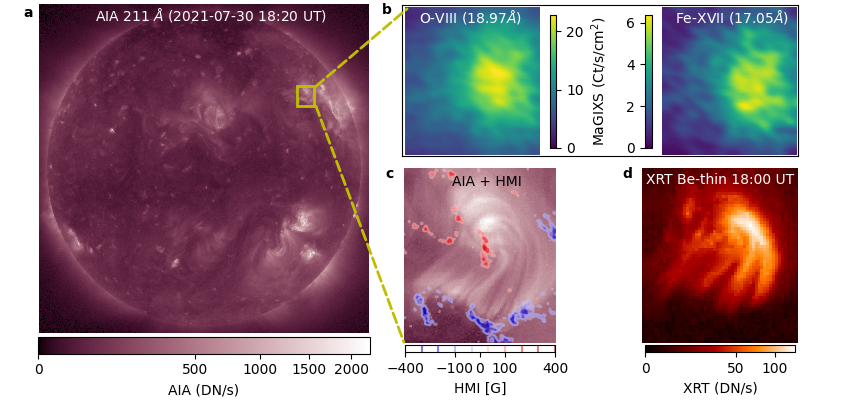}
\caption{Panel a shows the full-disk EUV image the observed by AIA 211 $\textup\AA$ passband, where the area of XBP-1 is marked by yellow box. Panel b shows spectrally pure maps of O-VIII and Fe-XVII spectrally pure images of XBP-1 derived from MaGIXS observations (Level 2 data product). Panel c shows the zoomed view of the yellow box shown in panel a. Red and blue contours represent the positive and negative polarities of observed line-of-sight photospheric magnetogram observed by HMI. Panel d shows the X-ray image of the XBP-1 observed by XRT.}
\label{fig-observation}
\end{figure*}

In the present study, we concentrated on the study of XBP-1 to understand its heating frequency by comparing the observations with simulated emissions from the hydrodynamic model. The location of the XBP-1 on the full disk image taken by the AIA/SDO 211 $\textup\AA$ passband is shown in Figure~\ref{fig-observation}a (yellow box).
{ Figure~\ref{fig-observation}b shows the spectrally pure maps of O-VIII and Fe-XVII at 18.97$\textup{\AA}$ and 17.05$\textup{\AA}$, derived from the MaGIXS observations as described in \citep{Sabrina_2023ApJ}. Each pixels of these images have a plate-scale of 2.8\arcsec$\times$2.8\arcsec, where as in \cite{Sabrina_2023ApJ}  the spectrally pure maps are shown with a plate-scale of 8.4\arcsec$\times$2.8\arcsec.}
Along with MaGIXS, we have used the concurrent observations of this XBP in EUV wavelength observed by AIA/SDO.
Level-1 AIA data were downloaded from Joint Science Operations Center (JSOC) and processed to level-1.5 using the standard procedure in SunPy~\citep{sunpy_community2020,sunpy_2022zndo}. 
We also used the synoptic X-ray images for this XBP observed by XRT/Hinode at the closest available time to the MaGIXS observation, which is 20 minutes before. 
Panels c and d show the zoomed view of XBP-1 observed by AIA 211$\textup\AA$ and XRT Be-thin passbands, respectively.
The red and blue contours in panel c represent the positive and negative polarities of the LOS photospheric magnetogram observed by SDO/HMI.
To model the XBP-1 emission, we need to know the coronal loop structures associated with this XBP. For this, we have extrapolated the observed photospheric magnetogram, as described in Section~\ref{sec-extrapolation}.

\section{Simulation of XBP-1}\label{sec-simulations}

XBP-1 is associated with bipolar magnetic field region, and consist of loop
structures as visible in EUV and X-ray images of AIA and XRT.
Most of the X-ray emission is associated with these loop-like structures.
Field-aligned hydrodynamic simulations are often used to simulate the emission of the plasma confined within these loops.
Here we have modeled the X-ray and EUV emission of the XBP-1 loops using the HYDRAD code. The HYDRAD code is described in detail in ~\cite{Bradshaw_2003A&A} and \cite{Bradshaw_2013ApJ...770...12B}. By accounting for the field-aligned gravitational acceleration and taking into account bulk transport, thermal conduction, viscous interactions, gravitational energy, Coulomb collisions, and optically thick radiation in the lower atmosphere transitioning to optically thin radiation in the overlying atmosphere, HYDRAD is able to solve the time-dependent equations for the evolution of mass, energy, and momentum for multi-fluid plasma (electrons and ions) in a given magnetic geometry.
HYDRAD can simulate the plasma response along the field-aligned direction for a given input heating profile and return the time evolution of temperature and density as a function of loop length. 
{ Here we employed the multi-species approach of HYDRAD, where electron and ions are treated as separate fluid. Also we consider the plasma is in equilibrium ionization and the loops have a constant  cross-section.}

We have derived the loop structures
associated with XBP-1 from the potential field extrapolation of the high resolution full-disk photospheric magnetograms observed by HMI/SDO as discussed in Section~\ref{sec-extrapolation}.  
These loops are simulated with HYDRAD using heating profiles that depend on the length and field strength as described in Section~\ref{sec:heating_function}.

\subsection{Magnetic field model}\label{sec-extrapolation}

\begin{figure*}[ht!]
\centering
\includegraphics[width=1\linewidth]{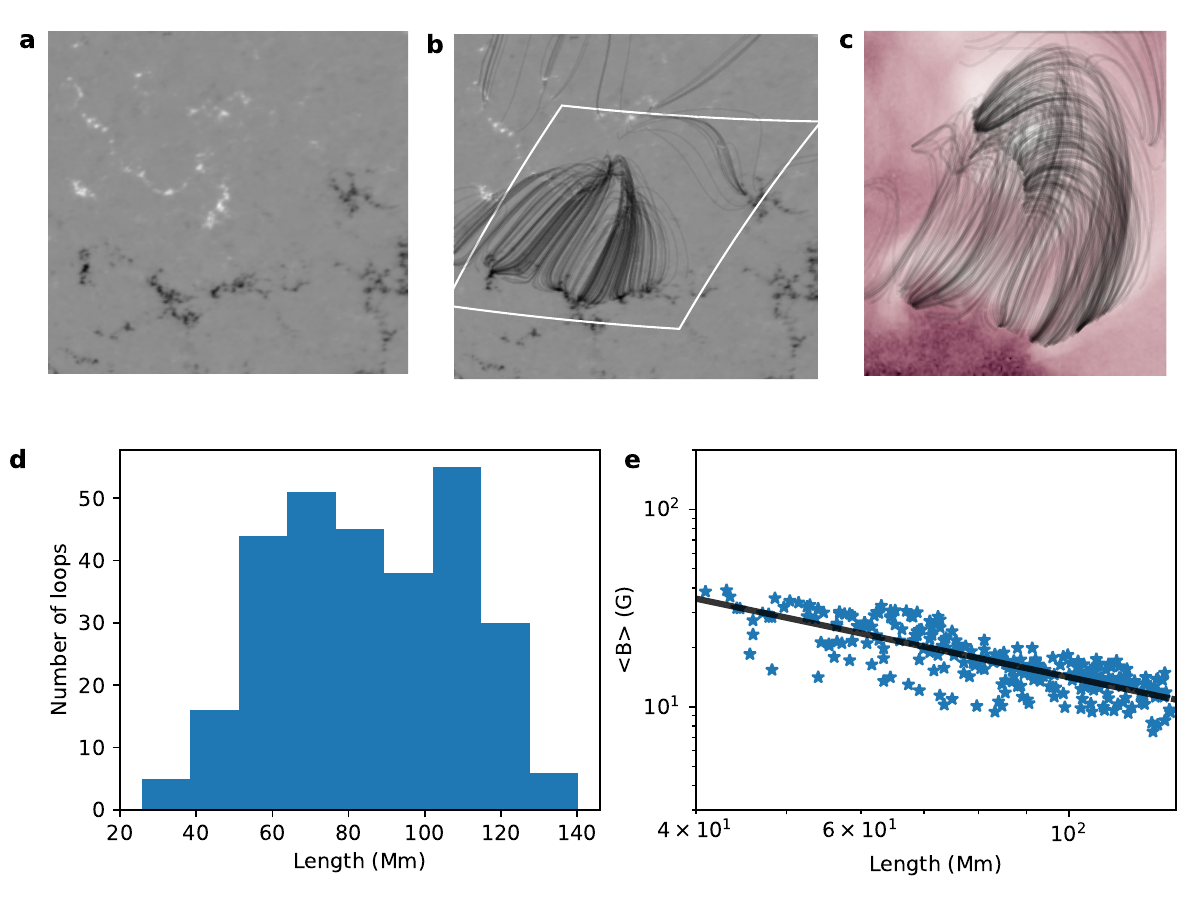}
\caption{Panel a shows the HMI magnetogram projected to an observers LOS of 50{\textdegree} latitude and 30{\textdegree} longitude. Extrapolated loops are overplotted on the magnetogram in panel b. The white box represents the FOV of XBP-1 as shown in Figure~\ref{fig-observation}. Panel c shows the projected extrapolated loops from Sun to Earth LOS, on-top of AIA 211 \textup{\AA} image. Panels d and e show the distribution of 300 extrapolated loops length and average magnetic field as a function of their lengths.}
\label{fig-extrapolation}
\end{figure*}

Using the locations of XBP-1 
we have identified its counterpart on the full-disk line-of-sight (LOS) HMI magnetogram, which is associated with a bipolar region. Considering these bi-poles as a lower boundary, 
we can extrapolate the field lines up to a height.
However, as this region is located away from the disk centre at a solar latitude and longitude of $\sim$50{\textdegree} and 30{\textdegree} respectively, the extrapolated loops might have a significant projection effect on the disk-plane. 
Thus using the \verb|reproject_to| functionality of SunPy \verb|Map| object, we have re-projected the HMI magnetogram to an observer line of sight at 50{\textdegree} latitude and 30{\textdegree} longitude.  Figure~\ref{fig-extrapolation}a shows the re-projected magnetogram.
From this magnetogram we have extrapolated the field lines up to a height of 500 HMI pixels ($\sim$180 Mm).
For this purpose, we have used the { Linear Force-Free (LLF) extrapolation code}, \verb|j_b_lff.pro|~\citep{nakagawa_1972,Seehafer_1978SoPh}, available within the SolarSoftWare package (SSW;~\citealp{Freeland_1998}). 
Using the three-dimensional extrapolated magnetic fields data, we have traced field lines through the volume corresponding to the XBP-1 following the streamline tracing method. 
For the streamline tracing we have chosen the seed points (through which field lines pass) randomly within the region of XBP-1 where absolute field strength is more than 20G at the base.
{ A force-free parameter, $\alpha$ $=$ -0.05 in our LLF model, provides a better match of the extrapolated loops with the observed emission in AIA passbands.} 
We have traced 300 loops, which is sufficient to represent the ensemble of the whole XBP-1 region in  visual inspections.
Projection of the extrapolated field lines on the magnetogram are shown in Figure~\ref{fig-extrapolation}b. Note that this field lines are projected with respect to a different observer's LOS than that of the AIA and XRT images shown in Figure~\ref{fig-observation}. Thus to compare with the morphology of the observed AIA and XRT images, we have re-oriented the 3D extrapolated loops towards Sun-Earth LOS by rotating it with same latitude (50\textdegree) and longitude (30\textdegree) and then take a projection as shown in Figure~\ref{fig-extrapolation}c. The loop morphology is now closely matches with the AIA and XRT images.
Figure~\ref{fig-extrapolation}d and \ref{fig-extrapolation}e show the distribution of all the extrapolated loop lengths and the length averaged magnetic field ($<B>$) distribution (see Equations 3 and 4 of \citealp{mondal_2023} for details). 
The loop length distribution has a peak near 100 Mm and the average magnetic field is found to vary inversely with loop length, similar to an AR as obtained by \cite{Mandrini_2000ApJ}.
The $<B> \propto L^{-1}$ relation is overplotted by a black dashed line as a reference.

\subsection{Heating profile}\label{sec:heating_function}

To simulate the coronal loops we assume  the loops are in hydrostatic equilibrium at the beginning (t=0 second) by setting a boundary conditions for foot points temperature and density. Once we provide the boundary values, HYDRAD calculates the initial temperature and density profile along the loops. 
We have chosen a footpoint temperature of 20,000 K, it seems reasonable to consider an isothermal chromosphere in the absence of any detailed knowledge~\citep{Bradshaw_2003A&A}.
The footpoint density is chosen such that the coronal loop average temperature remains at a value close to 0.5 MK (see Appendix-\ref{Append-initialConditions} for details), which is a reasonable lower boundary condition in the absence of any external heating.

We consider that loops are continuously heated by nanoflares~\citep{parker_1988, klimchuk_2015RSPTA}, those can occur with the release of stored magnetic energy that derives from slow photospheric driving. To derive the energy and occurrence frequency of these nanoflares, we employed the approach of \cite{mondal_2023}. Here we will briefly describe it.

We define the nanoflare heating in terms of a series of symmetric triangular profiles having a duration ($\tau$) of 100 s, similar to previous studies, e.g., \citealp{klimchuck_2008ApJ,cargill_2012b,Barnes_2016a}. The peak heating rate of an nanoflare is randomly chosen between the minimum ($H_0^{min}$) and  maximum ($H_0^{max}$) values associated with a loop.
The maximum energy density is considered to be equal to the stored magnetic energy due to the misalignment of the loop from vertical. If $\theta$ is the tilt of the magnetic field from the vertical, 
then the $H_0^{max}$ associated with i$^{th}$ loop would be,
\begin{equation}\label{eq:max_heating_rate}
    H^{max}_{0_{i}} = \frac{1}{\tau} \frac{(tan(\theta) <B>_{i})^2}{8\pi} (erg\hspace{0.1cm} cm^{-3}\hspace{0.1cm}s^{-1}) 
\end{equation}
We consider $H^{min}_{0}$ as one percent of $H^{max}_{0}$. Here, $\theta$ is known as Parker angle and it has been found that to satisfy the observed coronal heating energy requirement, the value of $tan(\theta)=c$, should be in the range of $0.2-0.3$~\citep{parker_1988,klimchuk_2015RSPTA}.

Because the free energy associated with a stressed loop is released during an impulsive event, releasing larger energy naturally creates a longer delay in accumulating enough energy to be released by the following event.  Taking this key consequence, we assume that the delay time between two successive events is proportional to the energy of the first event.
The delay time between $(l-1)^{th}$ and $l^{th}$ event will be,
\begin{equation}\label{eq-heating_delay} 
    d^l_{i} = \frac{\tau L}{F_i} \times H^{l-1}_{i}
\end{equation}
Here, $F_i$ is the Poynting flux associated with the i$^{th}$ loop by the photospheric driver.

\begin{figure}[ht!]
\centering
\includegraphics[width=1\linewidth]{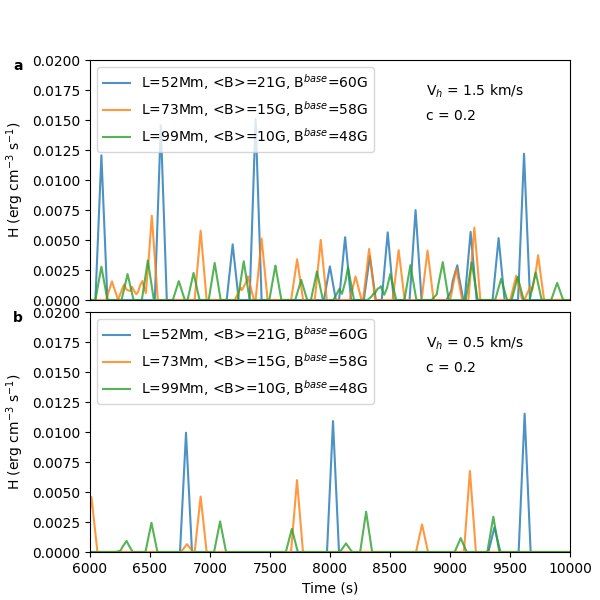}
\caption{Heating profile for three loops taken from the extrapolated distribution shown in Figure~\ref{fig-extrapolation}. Panel a and b corresponds to the photospheric driver velocity ($V_h$) 1.5 km/s and 0.5 km/s, respectively.}
\label{fig-heating_profile}
\end{figure}

\cite{mondal_2023} estimated $F$ by two different methods. In the first method they assumes that all the loops associated with all the XBPs have the same average Poynting flux, which is calculated from the observed DEM, and in the second method they consider a different Poynting flux for each loops derived from the expression of Poynting flux by the photospheric driver~\citep{Klimchuk_2006SoPh}. Here we have used the modified expression for the Poynting flux considering the scenario of expanding loops with coronal height (\citealp{mondal_2023}).

\begin{equation}\label{eq_poynFlux}
    F = -\frac{1}{4\pi} V_h tan(\theta) B^{base} <B>
\end{equation}
Here, V$_h$ is the horizontal speed of the flow that drives the field, $<B>$ is the average field strength along the loop, $B^{base}$ is the magnetic field at the coronal base. 
Figure~\ref{fig-heating_profile} show the estimated heating profile for three loops taken from the extrapolated distribution as shown in Figure~\ref{fig-extrapolation}. Panel a and b show the heating profile associated with V$_h$ = 1.5 km/s and V$_h$ = 0.5 km/s respectively with a similar value of c = 0.2.
Loops with larger length and lesser magnetic field strength, produces more higher frequency nanoflares compared with the loops with lower length and higher magnetic field strength. 

\begin{figure*}[ht!]
\centering
\includegraphics[width=1\linewidth]{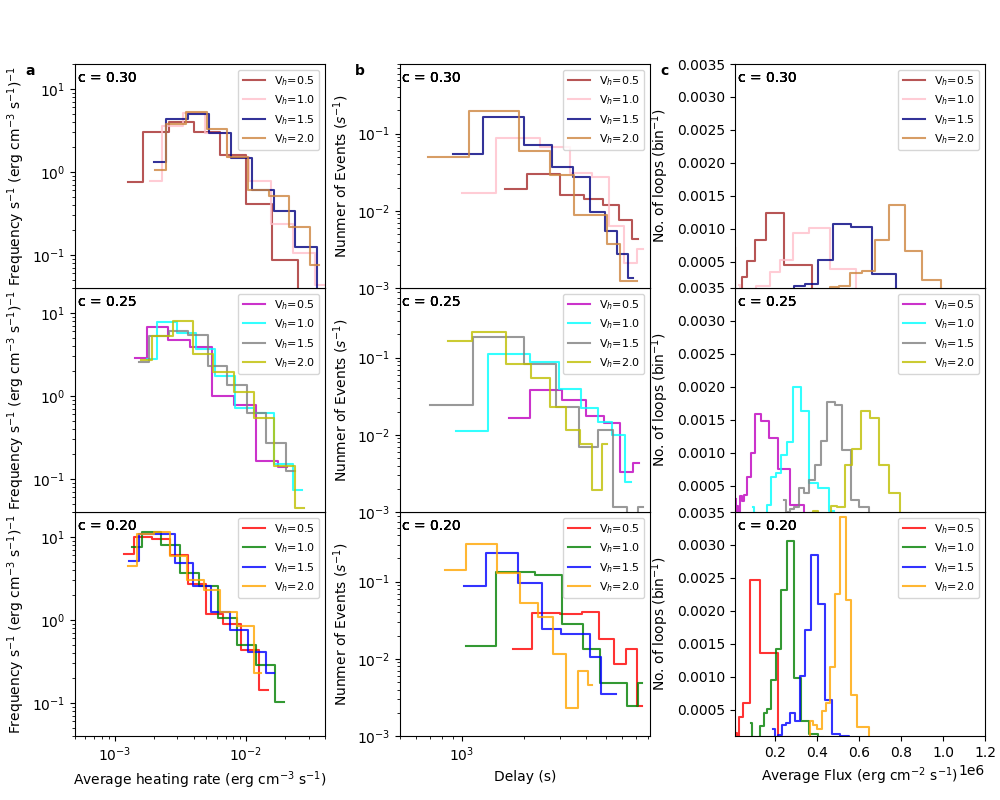}
\caption{Variation of the composite distribution of heating events for different heating parameters. Column a shows the average frequency distribution for all the loops as a function of heating rate. Column b shows the distribution of delay time between successive events for all the loops, where the average delay for each distribution are labeled. Column c shows the distribution of average Poynting flux for all the loops.}
\label{fig-heating_events}
\end{figure*}

\subsection{Simulation runs and outputs}\label{sec-simout}

Once we get the loop lengths and heating profiles for all the loops, 
we run the HYDRAD code for individual loops in a parallel computing environment using Pydrad\footnote{\url{https://github.com/rice-solar-physics/pydrad}}~\citep{pydrad_2023zndo} interface.
For each loop we consider 5 Mm as chromospheric height from each loop foot-points.
The location of each nanoflare event within a loop is determined from Poisson probability by considering expected location at the loop top with a significantly larger scale length (which determines how the heat will spread along the loop) in order to prevent localized heating.  
We simulate the evolution of the loops for the duration of 10,000 s and store the evolution of temperature and density for spatial grids of width 0.3 Mm (similar to HMI resolution) at a cadence of 25 s.
Using these temperatures and densities for the last 1000 s of evolution, we calculate the time averaged DEM for each grid points by considering a LOS plasma depth equal to the grid spacing ($dl$) along the loops. 
Note that assuming a LOS plasma depth equal to the grid spacing is a correct assumption for the $dl$ almost parallel to the observer LOS, but this may not be appropriate for the portions of the loops, mostly at higher height, where $dl$ is almost perpendicular to the observer LOS. 
{Here, we used a LOS depth of $\sim$0.3 Mm (=$dl$) for the DEM calculation. According to the HI-C observations (e.g, \citealp{Peter_2013A&A}) the smallest loops can have a diameter of 0.2 Mm, while the larger one could be 1.5 Mm.
Thus, depending on the actual loop diameter the loop-top emission might be slightly overpredict/underpredict. 
However, due to the lesser plasma density at loop-top, this would not affect the average DEM ($\propto$ $n^2 dl$) significantly.}
We create the DEM in the temperature range of logT = 5.6 to logT = 7.0 with $\delta$(logT) = 0.1.

Using the DEM values and the projected coordinates of the loop grids (Section~\ref{sec-extrapolation}) on the HMI magnetogram, we have created DEM maps for all the loops associated with XBP-1.
Folding this DEM map with the temperature response functions of different passbands of AIA, XRT, and MaGIXS, we generate the synthetic images of the XBP-1 associated with each passbands for the observation exposure time. Also we apply Poisson statistics to the pixel counts.
The AIA and XRT temperature responses ($R_i$) are generated using standard routine available in SSW package by applying passbands degradation correction at the time of observation.  
As our DEM maps are in the resolution of HMI plate-scales, the obtained $R_i$ for AIA and XRT are converted to HMI plate-scale. Also we have applied a cross-calibration factor of 2 for XRT responses as suggested by earlier studies~\citep{Schmelz_2015ApJ...806..232S,Wright_2017ApJ...844..132W,Athiray_2020}. The temperature responses of the different ions observed by MaGIXS are generated by multiplying the estimated contribution functions from CHIANTI~\citep{Dere_1997A&AS..chianti,chiantiV10_Zanna2020} with the MaGIXS effective area. For all the instruments we have used coronal abundances \citep{Feldman_1992b}.

As the simulated DEM maps are in HMI resolution, the synthetic images have the same HMI resolution of 0.5{\arcsec}.  
To compare the synthetic images with the observation, they are re-binned with the instrument plate-scale and then convolved with the instrument point spread function (PSF). 
{ Plate-scales of 0.6\arcsec, 2\arcsec, and 2.8\arcsec are used for AIA, XRT, and MaGIXS. We used the \verb|scipy.ndimage.gaussian_filter|~\citep{2020SciPy-NMeth} method for a Gaussian PSFs with FWHM of 1.2{\arcsec} for AIA, 2{\arcsec} for XRT, and 30{\arcsec} for MaGIXS (similar to the actual PSFs).}  

We repeat the simulation and create the synthetic images for different heating parameters ($c$ and V$_h$), which determine the heating profile. 
{ Depending on the observation and coronal energy losses, the values of $V_h$ and $c$ are chosen in the range of 0.5-2.0 km/s and 0.2-0.3 
~\citep{klimchuk_2015RSPTA}.}
Figure~\ref{fig-heating_events}a shows composite distribution of peak heating rates of nanoflares for all the loops associated with different combination of heating parameters. All the combination the heating rate frequency naturally follow a powerlaw with its heating rate. 
Figure~\ref{fig-heating_events}b shows the distribution of delay time between successive heating events for all the combination of heating parameters.
{ Combining Eq~\ref{eq-heating_delay}, \ref{eq:max_heating_rate}, and \ref{eq_poynFlux}, 
the delay time is proportional to $c$ and inversely proportional to $V_h$.}
{ Thus in Figure~\ref{fig-heating_events}b increasing $c$ represent more low-frequency events (larger delay time) compare to high-frequency events. On the other-hand increasing $V_h$ have more high-frequency (lower delay time) events compare to low-frequency events.}
Changing heating parameters will change the Poynting flux associated with the loops, which determines the effective heating. Distributions of the average Poynting flux associated with all the loops are shown in Figure~\ref{fig-heating_events}c.
{ The Poynting flux is proportional to $V_h$ and $c$ (Eq~\ref{eq_poynFlux}), causing increase in flux with the increase of either  $V_h$ or $c$.}

\section{Results and discussion}\label{sec-results}

\begin{figure*}[ht!]
\centering
\includegraphics[width=1\linewidth]{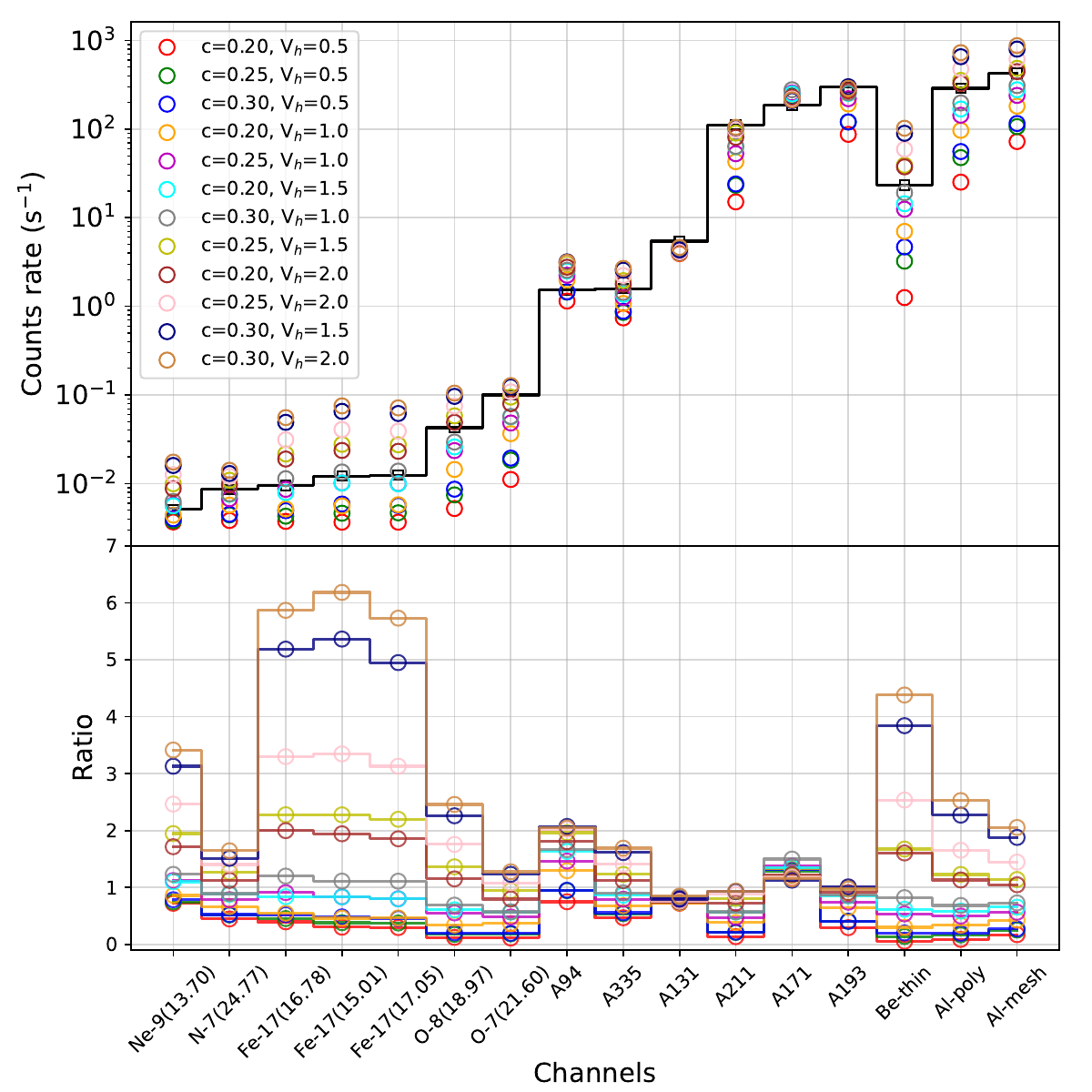}
\caption{Comparison between observed (black) and predicted (colored circle) average counts in all instruments passbands. The ratio between predicted and observed counts are shown in bottom panel.  Different color represent different heating parameters as mentioned in the label (c and V$_h$).}
\label{fig-com_sim_obs_int}
\end{figure*}

{ 
In this study we performed hydrodynamic simulation of an XBP, observed by MaGIXS to understand the nanoflare heating properties to maintain the heating of the XBP to the coronal temperature ($>$ 1 MK).
Simulations are runs for different combination of heating parameters, and for each of them, the AIA, MaGIXS, and XRT images are synthesized as described in Section~\ref{sec-simout}.
The spatially averaged intensities at different passbands of MaGIXS, AIA, and XRT are then compared with the average observed intensity.
Figure~\ref{fig-com_sim_obs_int} (top panel) shows the comparison for all combination of heating parameters as a function of instrument passbands plotted in abscissa. The black solid line represents the observed intensities, whereas the colored circles represent the synthetic intensities associated with different heating parameters. 
The absolute values of synthetic intensity could deviate from observed intensity due to various factors, e.g., the choice of number of loops associated with the XBP, but the intensities should be off by a consistent ratio for all the passbands. 
Thus our intention is not to compare the absolute values of the observed and predicted intensities; rather we compare the ratios between the predicted and the observed intensities for all the passbands.
The ratio of predicted to observed intensities is shown in bottom panel of Figure~\ref{fig-com_sim_obs_int}. 
}

\begin{figure}[ht!]
\centering
\includegraphics[width=1\linewidth]{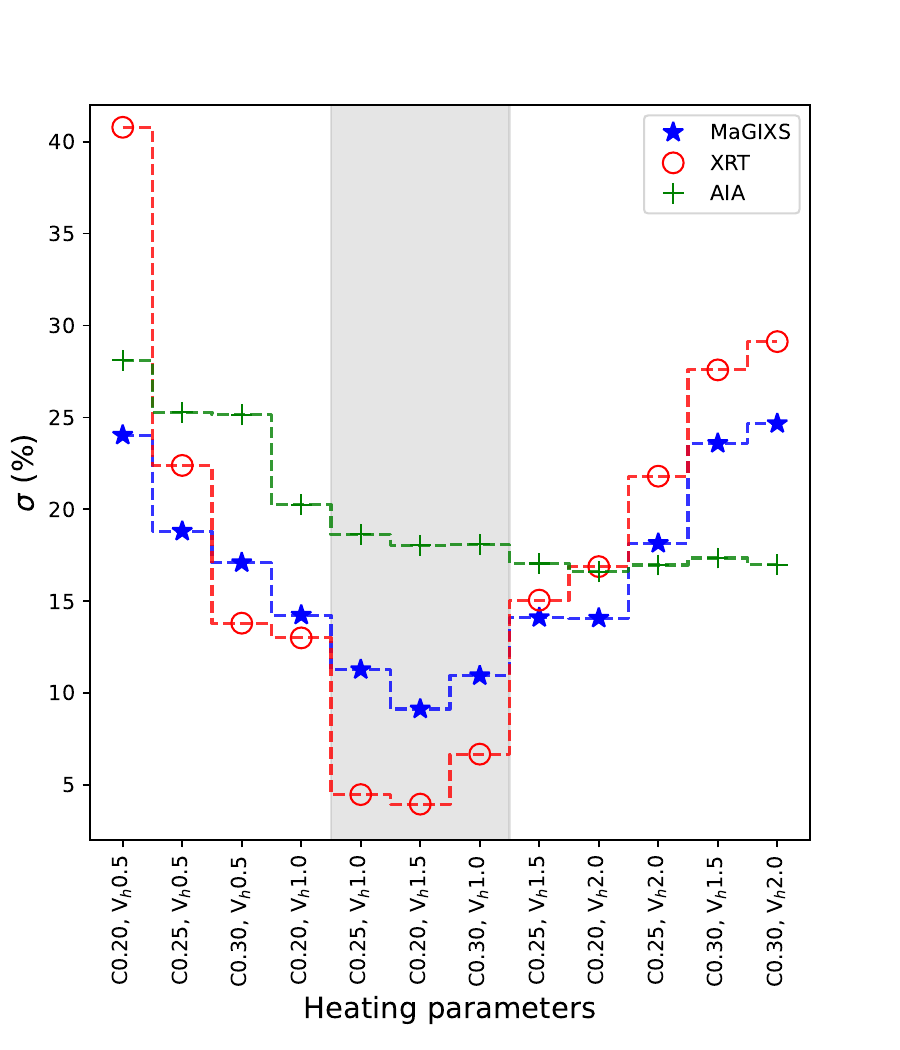}
\caption{Variation of $\sigma$ for different heating parameters, as mentioned in text. Grey shaded background shows the range of the heating parameters for which $\sigma$ converges to minimum values.}
\label{fig-sigma}
\end{figure}

We expect a similar ratio for all the passbands for a set of heating parameters, that can explain the observation.
To quantify how these ratios deviate from each other for a set of heating parameters, we derived the  standard deviation ($\sigma$) from their mean value. 
A smaller value of $\sigma$ indicates less deviation of the ratios from their mean and vice-versa. 
{ The comparison of $\sigma$ for different instrument passbands as discussed above would be more appropriate if cross-calibration factors among the instruments are well known, which is currently limited and being planned for the upcoming MaGIXS-2 flight\citep{Athiray2024A}.}
Therefore, here we compare the $\sigma$ for different instruments separately. Figure~\ref{fig-sigma} shows the $\sigma$ values for MaGIXS, AIA, and XRT. The Y-axis represent the $\sigma$ and X-axis shows the heating parameters.
We found that the $\sigma$ is converging (grey shaded region) towards a minimum value for both XRT and MaGIXS,  indicating a better match between predicted and observed intensities.
Whereas for AIA $\sigma$ is less variable for different heating parameters, indicating that using only the AIA passbands provides less sensitivity to determining the heating parameters and hence the heating frequency.
{ In the present study, most of the AIA passbands are sensitive to the cool/warm (around or below 1 MK) plasma, which is the reason for the AIA passbands are less sensitive to the heating parameters as discussed later in this section.}

{ Table~\ref{table-II} summarizes the values of heating parameters for which both MaGIXS and XRT show converging $\sigma$.
The range of the average Poynting flux associated with converging $\sigma$ is 3.0$\times$10$^5$ erg cm$^{-2}$ s$^{-1}$ to  4.0$\times$10$^5$ erg cm$^{-2}$ s$^{-1}$, which is similar to the average Poynting flux of coronal XBPs derived by \cite{mondal_2023} during the minimum of solar cycle 24.
This Poynting flux is more than one order of magnitude smaller than that of  ARs ($\sim$10$^7$ erg cm$^{-2}$ s$^{-1}$) as predicted by \cite{Withbroe_1977}, which is expected by considering the less magnetic activity of the XBPs.
}
\begin{deluxetable}{c c c c }
\tablecaption{Heating parameters which explains the observations.}
\label{table-II}
\tablehead{
c = $\tan(\theta)$  & V$_h$(km/s) & Average$_{delay}$ & Average$_{flux}$ \\
& km/s & s & 10$^5$ erg cm$^{-2}$ s$^{-1}$}
\startdata
0.25  & 1.0 & 2700 & 3.0 \\
0.20  & 1.5 & 1600 & 3.8 \\
0.30  & 1.0 & 2900 & 4.1 \\
\enddata
\end{deluxetable}

{ The average delay times between the nanoflares are in the range of 1500 s to 3000 s. These time range is smaller than the average cooling times (order of 10$^4$ s) of the  loops according to the formula given by \cite{cargill_2014ApJ...784...49C}, assuming similar equation parameters as those used by \cite{Barnes_2021ApJ...919..132B}.}
This suggests that the heating is dominated by high-frequency nanoflares, which is further supported by the fact that MaGIXS did not observe very hot ($>$ 5 MK) plasma for this XBP~\citep{Sabrina_2023ApJ}. 

\begin{figure*}[ht!]
\centering
\includegraphics[width=0.9\linewidth]{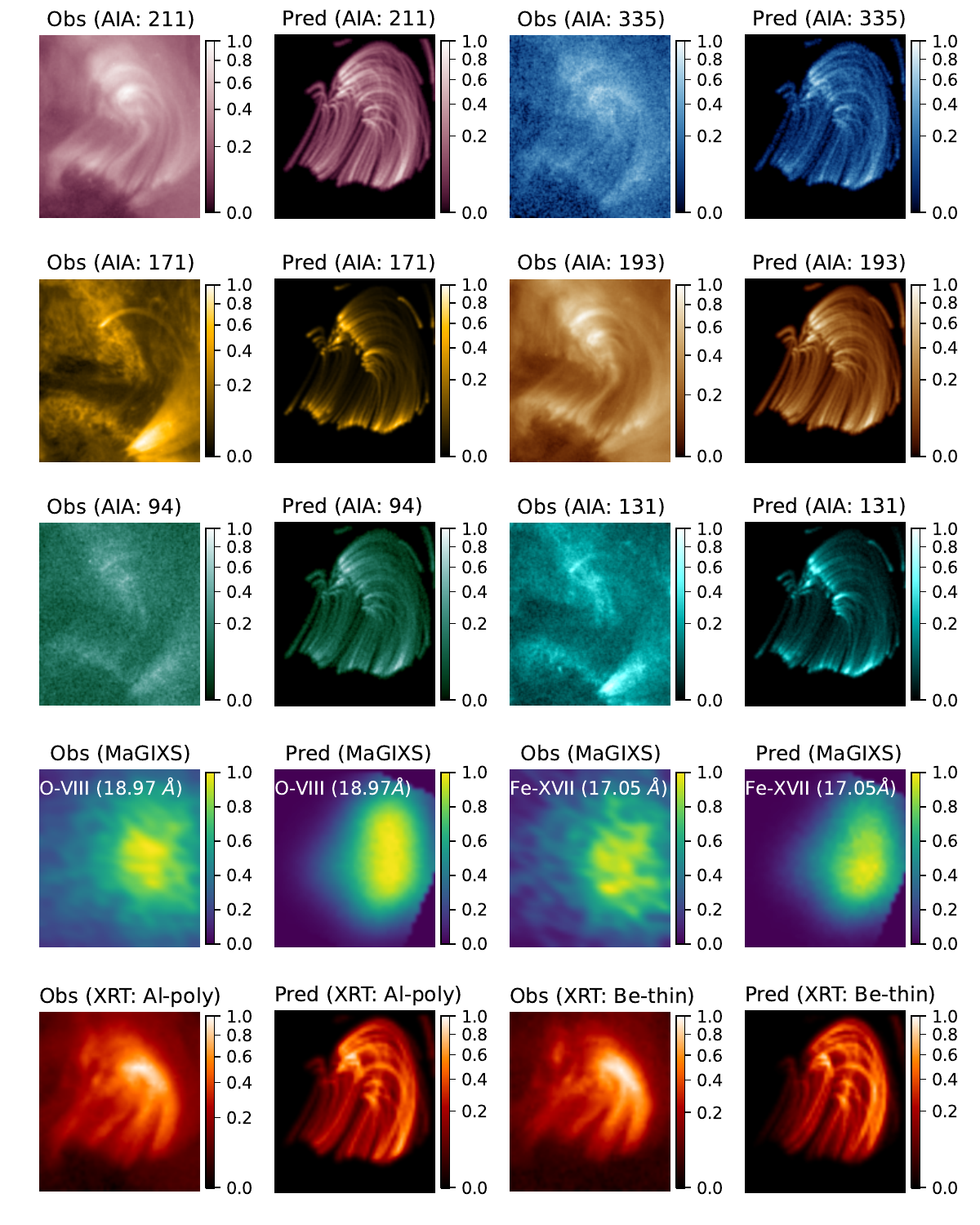}
\caption{Observed (columns: 1 and 3) and predicted (columns: 2 and 4) images { (color-bars are normalized with the maximum counts of each image)} in different passbands of AIA, MaGIXS, and XRT as mentioned in the label. The predicted images are for the model with heating parameters, c=0.2 and V$_h$=1.5 km/s. Note that the observed XRT images are 20 minutes before the predicted images.}
\label{fig-com_sim_obs_aia_xrt_img}
\end{figure*}

Figure~\ref{fig-com_sim_obs_aia_xrt_img} shows the representative comparison of observed and simulated images for  the heating parameters, $c=0.2$ and $V_h=1.5$ km/s in different passbands of AIA, MaGIXS and XRT. { As we are not comparing the absolute intensities, the images are normalized with their maximum pixel values.}
{ The overall emission morphology of the XBP in synthetic AIA images closely matched the observed images, except in a few places, such as the bright, cool (approximately 1 MK) structure in the bottom right of the observed 171$\textup\AA$ and 131$\textup\AA$ passbands. A closer inspection of this bright structure reveals its association with different sets of coronal loops that are not present in our magnetic model. Also, observed images has a diffuse background emissions, which is not present in our magnetic model and hence in the simulated emission.
Due to the poor spatial resolution of the MaGIXS, the loop structure of the XBP is not present in both observed and predicted images, but they show the brightening at similar locations. However, the synthetic O{\sc{xiii}} image shows a slightly elongated brightening in y-direction than the observed one.
The emission morphology in XRT synthetic images are slightly different than the observed emission. We think this might be due to the fact that the XRT observed images are 20 minutes earlier than the synthetic images, and at that time the magnetic field morphology was slightly different.}
In addition, we observed that the synthetic image in MaGIXS Fe{\sc{xviii}} passband does not predict a significant emission indicating the absence of hot ($>$ 5 MK) plasma, which is consistent with the observations as discussed by \cite{Sabrina_2023ApJ}. 
Also, the synthetic emission measure weighted temperature of the XBP for c=0.2 and V$_h$=1.5 km/s is found to be around 2 MK, which is similar to the predicted temperature from MaGIXS observations.

\begin{figure*}[ht!]
\centering
\includegraphics[width=1\linewidth]{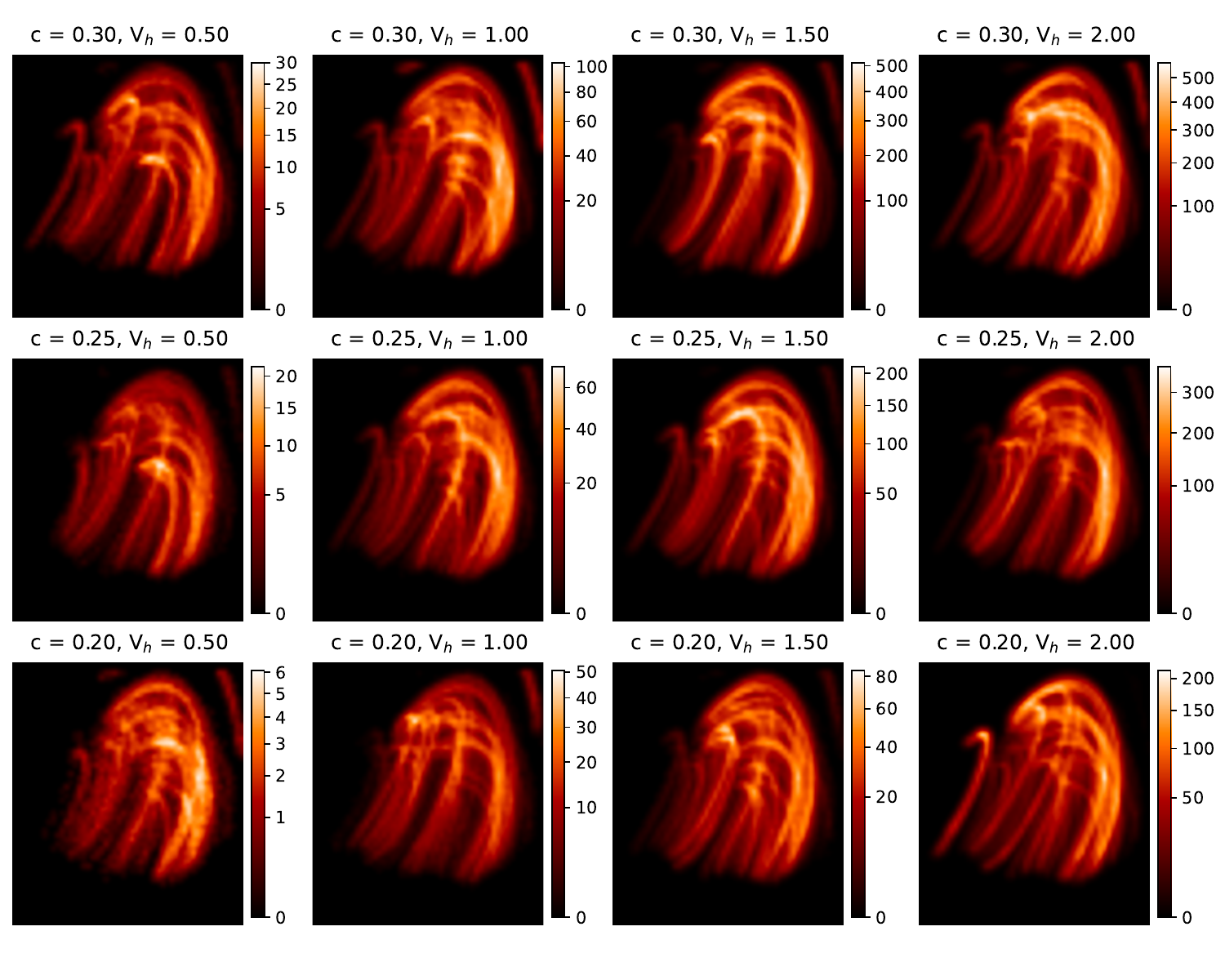}
\caption{{ Showing the variation of emission morphology in predicted emissions of XRT Be-thin passbands for all simulation runs associated with different heating parameters (mention at the top of each panel). The unit of the colorbar is DN/s.}}
\label{fig-sim_all_xrt_be_thin}
\end{figure*}

{ Figure~\ref{fig-sim_all_xrt_be_thin} showcase how the emission morphology varies with the heating parameters in XRT Be-thin filter. 
It is clearly visible that the synthetic emission morphology in XRT Be-thin filter is strongly dependent on the heating parameters and we found a similar results for the passbands sensitive to high temperature. This means that heating parameters are very sensitive to match high temperature emission.
A similar conclusion can be drawn by looking into the ratios of the spatially averaged synthetic and observed intensities as shown in the bottom panel of Figure~\ref{fig-com_sim_obs_int}. 
The ratios associated with different heating parameters show larger spread for the passbands that exhibit  high temperature sensitivity. For instance, Fe{\sc{xvii}} and Ne{\sc{ix}}, whose peak emissivity temperature (T$_{max}$) occur at log T = 6.75 and 6.6 show large spread ($\sim$0.5 to 6) in ratios; O{\sc{viii}} with T$_{max}$ at log T = 6.5 exhibit lesser spread ($\sim$0.5 to 2.5) in ratios; the cooler MaGIXS passbands O{\sc{vii}} and N{\sc{vii}} with T$_{max}$ at log T = 6.3 clearly show ratios varying $<$ 2. A similar trend is also clearly observed with AIA passbands. Having lesser sensitivity of the hot plasma in AIA passbands, they show a lesser variation in the intensity ratios.}

In this study we established a methodology to study the nanoflare heating frequency in coronal structures. This would be useful to study the capabilities and requirements of the upcoming instruments  to diagnose the heating frequency. 
As MaGIXS is sensitive to the hot component of the AR, in the upcoming flight of MaGIXS, if it will observe the hot ($>$ 5 MK) AR, where LF events are expected to occur, a similar analysis would be very useful to separate the contribution of HF and LF nanoflare for coronal heating budgets.

Our formulation of nanoflare heating profile in the present study is bashed on the dissipation of magnetic energy. In future a similar methodology could be useful to study the contribution of wave heating by adopting the wave heating scenario in the model, such as described by \cite{Reep_2018ApJ...853..101R}. 

\section{Summary}\label{sec-summary}

We have studied the nanoflare heating frequency of an XBP observed during the the first successful flight of MaGIXS along with { nearest available} observations by SDO/AIA, and Hinode/XRT. 
We compared the observed emission of this XBP with the simulated emission. 
The 1D hydrodynamic simulation code, HYDRAD, is used to simulate the XBP loops. The geometrical properties of the loops  are derived from the linear force-free field extrapolation of the observed photospheric magnetogram by SDO/HMI.
The loops are assumed to be heated by random nanoflare events depending on their length and magnetic field strength. 
Simulated emission in all instruments closely matches with the nanoflare heating model, with average Poynting flux in the range of { 3.0$\times$10$^5$ to 4.0$\times$10$^5$ erg cm$^{-2}$ s$^{-1}$.}
The average delay time between the nanoflares is found to be { 1500 s to 3000 s,} which is smaller than the average cooling time of the loops, suggesting the heating is dominated by high-frequency nanoflares.
Also, we have investigated the sensitivity of MaGIXS, XRT, and AIA passbands to diagnose nanoflare frequency.
We found that in our method, where we compare the average intensities of observed and synthetic images, the XRT and MaGIXS passbands are sensitive enough to diagnose the average nanoflare heating frequency, whereas AIA is the least sensitive. \\

\acknowledgments{
We acknowledge the Marshall Grazing Incidence X-ray Spectrometer (MaGIXS) instrument team for making the data available through the 2014 NASA Heliophysics Technology and Instrument Development for Science (HTIDS) Low Cost Access to Space (LCAS) program, funded via grant NNM15AA15C. MSFC/NASA led the mission with partners including the Smithsonian Astrophysical Observatory, the University of Central Lancashire, and the Massachusetts Institute of Technology.  MaGIXS was launched from the White Sands Missile Range on 2021 July 30.
BM's research was supported by an appointment to the NASA Postdoctoral Program at the NASA Marshall Space Flight Center, administered by Oak Ridge Associated Universities under contract with NASA. GDZ and HEM acknowledge the support of STFC. PT was supported for this work by NASA contract NNM07AB07C (Hinode/XRT) to the Smithsonian Astrophysical Observatory.
We also acknowledge the helpful comments from an anonymous
reviewer.
}

{\textit{Facilities:} SDO(AIA, HMI), Hinode (XRT), MaGIXS.}

{\textit{Software:} {Astropy~\citep{Astropy2018AJ....156..123A}, IPython~\citep{ipython_2007CSE.....9c..21P},matplotlib~\citep{matplotlib_2007CSE.....9...90H}, NumPy~\citep{2020NumPy-Array}, scipy~\citep{2020SciPy-NMeth},
SunPy~\citep{sunpy_community2020}, SolarSoftware~\citep{Freeland_1998}.}}

\renewcommand\thefigure{\thesection.\arabic{figure}}
\setcounter{figure}{0}

\appendix 

\section{Initial conditions}\label{Append-initialConditions}

The initial conditions are the initial temperature and density profiles along the loop length at t=0 s, that HYDRAD evolve with time subject to some external driver. More information on configuring the initial conditions can be found in the HYDRAD user manual
\footnote{\url{https://github.com/rice-solar-physics/HYDRAD/blob/master/HYDRAD_User_Guide(03_20_2021).pdf}}.
At the beginning, we consider the loops are in hydrostatic equilibrium, which ensures that at later time (t $>$ 0 s) the evolution is only due to the external driver.
Taking into account a few simplified assumptions, in hydrostatic equilibrium (see Equation-9 in \citealp{reale_2014}), we can write,
\begin{equation}\label{eq-1}
    P \approx \frac{1}{L}(\frac{T_{max}}{1.4\times 10^3})^3
\end{equation}
Here, P is the uniform pressure throughout the loop of length L and T$_{max}$ is the maximum temperature. 

We want to keep the loop (above chromosphere) with an average temperature (T$_{avg}$), e.g., 0.5 MK, which is a reasonably lower value.
Following, \cite{Cargill_2012} we can write,
\begin{equation}
    T_{avg} \approx 0.9\times T_{max}
\end{equation}
Also from ideal gas law,
\begin{equation}\label{eq-3}
    P = nkT
\end{equation}
Combining Eq.\ref{eq-1}-\ref{eq-3}, foot-point density would be,
\begin{equation}\label{eq-4}
    n_{base} = \frac{P_{base}}{k T_{base}}
\end{equation}

Following \cite{Bradshaw_2003A&A} a footpoint temperature of 20,000 K is physically reasonable to treat a stratified, isothermal chromosphere (where the scale height is constant) in the absence of a detailed knowledge and thorough treatment. 
Note that, the chromospheric density is important; if it is too low then a very strong nanoflare could essentially ablate the entire mass content of the chromosphere into the corona, emptying it out, and causing the transition region (basically, a thermal conduction front) to hit the edge of the computational domain, which is not desirable. A denser chromosphere can be obtained by choosing a lower isothermal temperature (e.g. 10,000 K instead of 20,000 K).

\begin{figure}[ht!]
\centering
\includegraphics[width=0.8\linewidth]{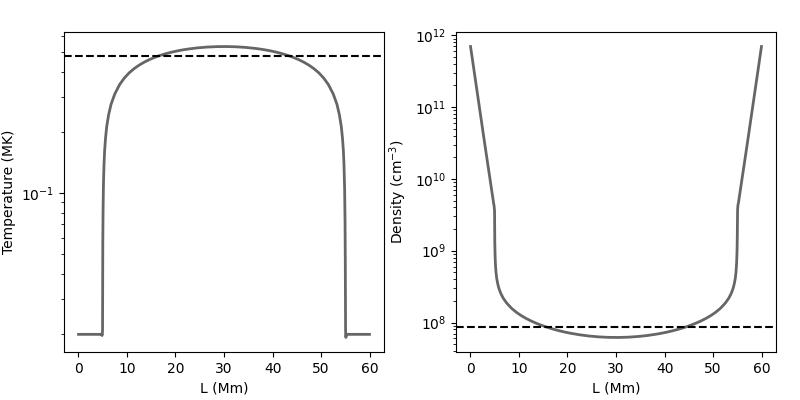}
\caption{Initial temperature (solid curve in left) and density (solid curve in right) profile solved by HYDRAD for a given loop footpoint temperature and density of 20,000  K and $4\times10^9$ cm$^{-3}$ respectively. The dashed horizontal lines are the loop averaged temperature and density. A height of 5 Mm for each end of the loop is considered as chromosphere.}
\label{fig-initial_cond}
\end{figure}

Consider, L = 60 Mm. Then from Eq.\ref{eq-4}, to maintain an average temperature of 0.5 MK throughout the coronal portion of the loop, the footpoint density, n$_{base}\approx$ $4\times10^9$ cm$^{-3}$.
Once we know the footpoint density and temperature and provide them to HYDRAD, the code will calculate the initial temperature and density profile along the loop, as shown in Figure~\ref{fig-initial_cond}. 

\newpage
\bibliography{myref}   
\bibliographystyle{aasjournal}

\end{document}